\documentclass[letterpaper, 10 pt, conference]{ieeeconf}  % Comment this line out if you need a4paper

\IEEEoverridecommandlockouts                              % This command is only needed if 
\usepackage[pdftex]{graphicx}
\usepackage{algorithm}
\usepackage{algpseudocode}

\usepackage{amssymb,amsmath, amsthm}
\usepackage{subfig}
\usepackage{indentfirst}
\usepackage{color}
\usepackage{comment}
\usepackage[hidelinks]{hyperref}
\usepackage{enumerate}
\usepackage{tikz}
\usetikzlibrary{shapes,arrows}
\usepackage{epstopdf}
\usepackage{epsfig}
\usepackage{wrapfig}
\usepackage[font=small]{caption}
\usepackage{balance}
\usepackage{setspace}

\def\mf{\mathbf}
\def\mb{\mathbb}
\def\mc{\mathcal}
\def\beq{\begin{equation*}}
\def\eeq{\end{equation*}}
\def\bql{\begin{equation}}
\def\eql{\end{equation}}
\def\bqn{\begin{eqnarray*}}
\def\eqn{\end{eqnarray*}}
\def\bnl{\begin{eqnarray}}
\def\enl{\end{eqnarray}}
\def\bna{\bql\begin{array}{rcl}}
\def\ena{\end{array}\eql}
\def\bnn{\beq\begin{array}{rcl}}
\def\enn{\end{array}\eeq}
\def\bma{\begin{bmatrix}}
\def\ema{\end{bmatrix}}
\def\bmx{\begin{matrix}}
\def\emx{\end{matrix}}
\def\ben{\begin{enumerate}}
\def\een{\end{enumerate}}
\def\bit{\begin{itemize}}
\def\eit{\end{itemize}}
\def\bei{\begin{itemize}}
\def\eei{\end{itemize}}
\def\bet{\begin{tabular}}
\def\eet{\end{tabular}}
\def\span{\text{span}}

\newcommand{\allcaps}[1]{\uppercase\expandafter{#1}}

\theoremstyle{definition}

\newtheorem{assumption}{Assumption}  
\newtheorem{theorem}{Theorem}

\newtheorem{lemma}{Lemma}
\newtheorem{remark}{Remark}

\makeatletter
\let\OldStatex\Statex
\renewcommand{\Statex}[1][3]{%
  \setlength\@tempdima{\algorithmicindent}%
  \OldStatex\hskip\dimexpr#1\@tempdima\relax}
\makeatother

\allowdisplaybreaks
\title{\LARGE \bf
 Temporally-Consistent Bilinearly Recurrent Autoencoders for Control Systems

}

\author{Ananda Chakrabarti$^{1*}$, Indranil Nayak$^{2*}$, Debdipta Goswami$^{3}$% <-this % stops a space
\thanks{This work is supported by OSU Departmental Startup fund.}
\thanks{$^{1}$Ananda Chakrabarti is a PhD student in the Department of Electrical and Computer Engineering, The Ohio State University, Columbus, OH 43210.
        {\tt\small chakrabarti.44@osu.edu}}%   
\thanks{$^{2}$Indranil Nayak is a PhD Candidate in the Department of Electrical and Computer Engineering, The Ohio State University, Columbus, OH 43210.
        {\tt\small nayak.77@osu.edu}}% 
\thanks{$^{3}$Debdipta Goswami is an Assistant Professor in the Department of Mechanical and Aerospace Engineering, The Ohio State University, Columbus, OH 43210.
        {\tt\small  goswami.78@osu.edu}}%
\thanks{$^{*}$First two authors contributed equally to this manuscript.}%
}

\begin{document}
%\onecolumn
\maketitle

\tikzstyle{block} = [draw, fill=blue!20, rectangle, 
    minimum height=3em, minimum width=4em]
\tikzstyle{sum} = [draw, fill=blue!20, circle, node distance=1cm]
\tikzstyle{input} = [coordinate]
\tikzstyle{noise} = [coordinate]
\tikzstyle{output} = [coordinate]
\tikzstyle{pinstyle} = [pin edge={to-,thin,black}]

\thispagestyle{empty}
\pagestyle{empty}

%%%%%%%%%%%%%%%%%%%%%%%%%%%%%%%%%%%%%%%%%%%%%%%%%%%%%%%%%%%%%%%%%%%%%%%%%%%%%%%%
\begin{abstract}
This paper introduces the temporally-consistent bilinearly recurrent autoencoder (tcBLRAN), a Koopman operator based neural network architecture for modeling a control-affine nonlinear control system. The proposed method extends traditional Koopman autoencoders (KAE) by incorporating bilinear recurrent dynamics that are consistent across predictions, enabling accurate long-term forecasting for control-affine systems. This overcomes the roadblock that KAEs face when encountered with limited and noisy training datasets, resulting in a lack of generalizability due to inconsistency in training data. Through a blend of deep learning and dynamical systems theory, tcBLRAN demonstrates superior performance in capturing complex behaviors and control systems dynamics, providing a superior data-driven modeling technique for control systems and outperforming the state-of-the-art Koopman bilinear form (KBF) learned by autoencoder networks.

\end{abstract}

%%%%%%%%%%%%%%%%%%%%%%%%%%%%%%%%%%%%%%%%%%%%%%%%%%%%%%%%%%%%%%%%%%%%%%%%%%%%%%%%
\section{INTRODUCTION}

Accurate modeling and prediction of nonlinear dynamical systems with inputs represents a cornerstone challenge in control engineering. Traditional physics-based modeling is becoming increasingly challenging for complex nonlinear systems with insufficient system-level knowledge. On the other hand, recent developments in machine-learning (ML) techniques for modeling complex systems from data have become useful in a wide variety of problems, e.g., classification, speech recognition \cite{Hinton2012}, board games \cite{Silver2016}, and even discovering mathematical algorithms \cite{Fawzi2022}. However, any knowledge of the underlying dynamical system provides valuable insights to generate appropriate features and interpretability to any ML algorithm, giving rise to a family of physics constrained learning (PCL) methods that incorporate constraints arising from physical consistency of the governing dynamical system. In this context, the Koopman operator theory \cite{Koopman1931, Rowley2009} emerges as a powerful tool, offering a linear perspective on inherently nonlinear dynamical systems. This theory enables the application of linear analysis techniques to nonlinear systems by operating in an infinite-dimensional function space, thereby linearizing the dynamics of observable functions. However, since the Koopman operator generally maps between infinite-dimensional function spaces, it cannot be represented computationally without a finite-dimensional projection. Early methods e.g., extended dynamic mode decomposition (EDMD) \cite{Williams2015} use large amount of data to approximate the Koopman operator and the latent linear dynamics by projecting it on carefully crafted dictionary functions. But the goodness of the approximation depends on the Koopman-invariance property of the dictionary. More recently, a class of ML algorithms has been developed to automate the choice of dictionary by using deep neural networks (DNNs) to find an approximate finite-dimensional Koopman invariant subspace. These Koopman autoencoder (KAE) or linearly recurrent autoencoder (LRAN) algorithms use an autoencoder network that maps the state-space into a latent space where the dynamics can be linearly approximated by the finite-dimensional Koopman operator encoded via a linear layer \cite{otto2019linearly, lusch2018deep}.

Despite the promising foundation laid by Koopman operator theory, its direct application to control systems poses significant challenges since the linear latent structure is no more valid with a time-varying control input. For a control-affine system, \cite{Goswami2021, Otto2020} proves the latent dynamics to be bilinear in the Koopman invariant subspace, thereby yielding Koopman bilinear form (KBF). The KAE/LRAN framework is applied to learn KBF for robotic system control \cite{Folkestad}, but the window for accurate prediction remains short with control-input. In a recent work by the authors, temporally-consistent Koopman autoencoder (tcKAE) \cite{Nayak2024} is developed for consistent and accurate prediction of \emph{autonomous} dynamical systems using a temporal consistency constraint with Koopman autoencoder framework. It compares the predictions from different initial time-instances to a final time in the latent space in contrast with prior KAE methodologies, where the enforcement of multi-step look-ahead prediction loss is reliant on \emph{labelled} data and latent space representation of the labelled data.  %Recent advancements in machine learning, particularly autoencoders, have facilitated the approximation of the Koopman operator, leading to the development of Koopman Autoencoders (KAEs). These models have shown potential in capturing the complex dynamics of autonomous systems. However, their adaptation to control-affine systems, which necessitates handling bilinear interactions between state variables and control inputs, remains an underexplored area.

This paper presents the temporally-consistent bilinearly recurrent autoencoders (tcBLRAN) for \emph{control} systems by employing the temporal consistency regularization developed in \cite{Nayak2024} to learn the KBF using a bilinearly recurrent autoencoder. The contributions of this manuscript are: (1) extension of traditional KAE/LRAN to control-affine nonlinear systems with bilinear recurrent autoencoders, thereby enforcing the Koopman invariance conditions developed in \cite{Goswami2021}; (2)  introduction of a prediction consistency mechanism that ensures the temporal coherence of model predictions, significantly enhancing long-term forecasting accuracy; (3) provide a theoretical justification to the prediction consistency from the time-invariance property of the control system; (4) comprehensive evaluation of the model-prediction performance of tcBLRAN and vanilla BLRAN for different control-affine nonlinear systems.
%traditional KAEs through:
% \begin{enumerate}
%     \item A novel architecture that seamlessly integrates control inputs with state variables, enabling the effective modeling of control-affine systems in a Koopman invariant subspace.
%     \item The introduction of a prediction consistency mechanism that ensures the temporal coherence of model predictions, significantly enhancing long-term forecasting accuracy.
%     \item A comprehensive evaluation against state-of-the-art methodologies, demonstrating tcBRKA's superior performance in modeling and predicting the behavior of control systems.
% \end{enumerate}

By bridging the gap between KAE and control-affine systems, tcBLRAN not only broadens the applicability of Koopman-based models but also sets a new benchmark in the predictive modeling of dynamic systems. Our contributions pave the way for future advancements in control theory, predictive modeling, and beyond, promising significant implications for a wide array of applications ranging from autonomous vehicles to energy management systems.

The paper is organized as follows: Section II provides an overview of Koopman spectral theory, describes Koopman bilinear form, and introduces Koopman autoencoders for dynamics learning. Section III proposes the tcBLRAN algorithm and provides its theoretical basis. Section IV illustrates the prediction performance with numerical examples. Section V summarizes the paper and suggests possible future work.

\section{Mathematical Preliminaries}
\subsection{Koopman theory: an overview}
Consider a dynamical system evolving on a compact manifold $\mb{X}\subseteq\mb{R}^d$,
\bql \label{Eq: Autonomous}
\dot{\mf{x}}=\mf{f}(\mf{x}),
\eql
where $\mf{x}\in\mb{X}$ and $\mf{f}:\mb{X}\rightarrow\mb{X}$. Let $\mf{\Phi}(t,\mf{x})$ be the flow map of the system (\ref{Eq: Autonomous}) at time $t>0$ starting from an initial state $\mf{x}$. A measurable function $\varphi:\mb{X}\rightarrow\mb{C}$ is called an observable of the dynamical system \eqref{Eq: Autonomous}. Let $\mc{F}$ be the space of all complex-valued observables $\varphi:\mb{X}\rightarrow\mb{C}$. The continuous-time Koopman operator is defined as $\mc{K}^t:\mc{F}\rightarrow\mc{F}$ such that
\bql
(\mc{K}^t\varphi)(\cdot)=\varphi\circ\mf{\Phi}(t,\cdot),
\eql
where $\circ$ denotes the function composition. Unlike the original system, the Koopman operator is linear over its arguments, i.e., observable functions, and therefore can be characterized by its eigenvalues and eigenfunctions. A function $\phi:\mb{X}\rightarrow\mb{C}$ is an eigenfunction of $\mc{K}^t$ if 
 \bql
 (\mc{K}^t\phi)(\cdot)=e^{\lambda t}\phi(\cdot),
 \eql
 with eigenvalue $\lambda\in\mb{C}$. The infinitesimal generator of $\mc{K}^t$, i.e., $\lim_{t\rightarrow 0}\dfrac{\mc{K}^t-I}{t}$, is $\mf{f}\cdot\nabla=L_{\mf{f}}$ \cite{Mauroy2016}, where $I$ is the identity operator and $L_{\mf{f}}$ is the Lie derivative with respect to $\mf{f}$. The infinitesimal generator satisfies the eigenvalue equation
 \bql \label{Eq: Eigenvalue}
 L_{\mf{f}}\phi=\lambda\phi.
 \eql
 Hence, the time-varying observable $\psi(t,\mf{x})\triangleq \mc{K}^t\varphi(\mf{x})$ is the solution of the partial differential equation (PDE) \cite{Mauroy2016}
 \bna \label{Eq: Observation PDE}
 \dfrac{\partial\psi}{\partial t}&=&L_{\mf{f}}\psi,\\
 \psi(0,\mf{x})&=& \varphi(\mf{x}).
 \ena
 \par In spite of its linearity, the Koopman operator is infinite-dimensional and has an infinite number of eigenfunctions. In fact, if $\phi_1$ and $\phi_2$ are eigenfunctions of $\mc{K}^t$ with eigenvalues $\lambda_1$ and $\lambda_2$, respectively, then $\phi_1^{k}\phi_2^l$ is also an eigenfunction with eigenvalue $k\lambda_1 + l\lambda_2$ for any $k,l\in\mb{N}$. Moreover, the Koopman operator, being infinite-dimensional, may contain continuous and residual spectra with a generalized eigendistribution \cite{Mohr2014}. However, the discussions in this paper are restricted to the point spectra of the Koopman operator.

Let $\mf{g}(\cdot)\in\mc{F}^p$, $p\in\mb{N}$, be a vector-valued observable. The observable $\mf{g}$ can be expressed in terms of Koopman eigenfunctions $\phi_i(\cdot)$ as follows:
 \bql
 \mf{g}(\cdot)=\sum_{i=1}^{\infty}\phi_i(\cdot)\mf{v}^{\mf{g}}_i,
 \eql
 where $\mf{v}^{\mf{g}}_i\in\mb{R}^p, i=1,2,\ldots,$ are called the \emph{Koopman modes} of the observable $\mf{g}(\cdot)$. Koopman modes form the projection of the observable on the span of Koopman eigenfunctions \cite{Budisic2012}. The Koopman eigenvalues and eigenfunctions are properties of the dynamics only, whereas the Koopman modes depend on the observable.

\subsection{Koopman bilinear form: Koopman approximation of control affine system}
Consider the control affine system \bna \label{Eq: System}
\dot{\mf{x}} &=& \mf{f}(\mf{x})+\sum\limits_{i=1}^m\mf{g}_i(\mf{x})u_i\\
\mf{y} &=& \mf{h}(\mf{x}),\quad \mf{x}_{t_0} = \mf{x}_0,
\ena
where $\mf{x}\in\mb{X}\subseteq\mb{R}^d$, $u_i\in\mb{R}$, for $i=1,\ldots,m$ and $\mf{y}\in\mb{R}^p$. Here the control input enters only through the control vector fields $\mf{g}_i$. Applying (\ref{Eq: Observation PDE}) to the system (\ref{Eq: System}), the evolution of the PDE, derived in \cite{Goswami2017}, is
\bna \label{Eq: Bilinear_Operator}
 \dfrac{\partial\psi}{\partial t}&=&L_{\mf{f}}\psi + \sum\limits_{i=1}^{m}u_iL_{\mf{g}_i}\psi,\\
 \psi(0,\mf{x})&=& \varphi(\mf{x}),
\ena
where $L_{\mf{g}_i}\triangleq\mf{g}_i\cdot\nabla,\,i=1,\ldots,m$ are the corresponding Lie derivatives and hence are linear operators on the space of $\psi$. The resultant bilinear forced PDE system \eqref{Eq: Bilinear_Operator} provides a way to represent the control-affine system \eqref{Eq: System} in a lifted latent space as a bilinear ODE. The following theorem from \cite{Goswami2017, Goswami2021} provides the sufficient condition for bilinearization of \eqref{Eq: System}:
\begin{theorem}
\cite{Goswami2017, Goswami2021} The system \eqref{Eq: System} is bilinearizable in a countable (possibly infinite) basis if the eigenspace of $L_{\mf{f}}$, i.e., the Koopman generator corresponding to the drift vector field is an invariant subspace of $L_{\mf{g}_i}, i=1,\ldots,m$, i.e., the Koopman generators related to the control vector fields.
\end{theorem}
\begin{remark}
This result is further extended in \cite{Otto2020} for any invariant subspace of $L_{\mf{f}}$, i.e., any Koopman-invariant subspace associated with the drift vector field that is invariant under $L_{\mf{g}_i}, i=1,\ldots,m$, thereby removing the necessity to find Koopman eigenfunctions.
\end{remark}

References \cite{Goswami2021}, \cite{Otto2020} further demonstrates that if a transformation $\mf{\Psi}(\cdot) \in \mc{F}^{N_l},\quad \mf{\Psi}(\cdot) = [\psi_1(\cdot),\ldots, \psi_{N_l}(\cdot)]^T$ can be found such that $\operatorname{span}\{\psi_1(\cdot),\ldots, \psi_{N_l}(\cdot)\}$ is invariant under $L_{\mf{f}}$ and $L_{\mf{g}_i}, i=1,\ldots,m$, then the transformation $\mf{z} = \mf{\Psi}(\mf{x})$ allows us to express the system \eqref{Eq: System} in the following Koopman bilinear form (KBF)
\bql \label{Eq: Bilinear}
\dot{\mf{z}} = A\mf{z} + \sum\limits_{i=1}^m B_i\mf{z}u_i,
\eql
where $A$ and $B_i$'s are the approximation of Koopman generators for drift and control-vector fields.

%The citation \cite{Goswami2021} goes like this.

% To put page spanning subfigures. Change to figure instead of figure* to keep it within a column.
% \begin{figure*}[t]
% \centering 
% \subfloat[]{\includegraphics[trim=0cm 0cm 0cm 0cm, clip=true, width=0.31\textwidth]{Fig1.png}}
% \subfloat[]{\includegraphics[trim=0cm 0cm 0cm 0cm, clip=true, width=0.35\textwidth]{Fig2.png}}
% \subfloat[]{\includegraphics[trim=0cm 0cm 0cm 0cm, clip=true, width=0.34\textwidth]{Fig3.png}}
% \caption{Caption: (a) (b) (c)}
% \end{figure*}

\subsection{Enforcing invariance via neural-networks: Koopman autoencoders}
Discovering a finite-dimensional Koopman-invariant subspace purely from data is a challenging task, and an active area of research. Recent works \cite{takeishi2017learning, lusch2018deep, otto2019linearly} leverage the nonlinear function approximation capabilities of neural networks to find a suitable transformation from the state space $\mb{X}$ to a Koopman invariant subspace via latent state $\mf{z}$. The basic operation of KAE can be decomposed into three components \eqref{eq:KAE_operation}, \textit{i)} an encoding $(\mf{\Psi}_e(\cdot)):\mb{X} \rightarrow \mb{R}^{N_l}$ that transforms the original state $\mf{x}$ to a Koopman observable $\mathbf{z}\in\mb{R}^{N_l}$, \textit{ii)} advancing the dynamics in that transformed space through a linear operator $(K^{\Delta t}\in\mb{R}^{N_l\times N_l})$, and \textit{iii)} a decoding $(\mf{\Psi}_d(\cdot))$ back to the original state space:
\bnl\label{eq:KAE_operation}
    \mathbf{z}_t\approx \mf{\Psi}_e(\mathbf{x}_t) & \rightarrow & \mathbf{z}_{t+\Delta t}=K^{\Delta t}\cdot \mathbf{z}_t \\\nonumber & \rightarrow & \mf{x}_{t + \Delta t} \approx \mf{\Psi}_d(\mf{z}_{t + \Delta t})=\hat{\mf{x}}_{t + \Delta t},   
\enl where $K^{\Delta t}$ is a finite dimensional, discrete-time representation of $\mc{K}^{t}.$ This can be extended to control-affine systems by leveraging the KBF and encoding the invariant subspace of $L_{\mf{f}}$ and $L_{\mf{g}_i}$s as 

\bnl\label{eq:Control_KAE_operation}
    \mathbf{z}_t\approx \mf{\Psi}_e(\mathbf{x}_t) & \rightarrow & \mathbf{z}_{t+\Delta t}= e^{(A +  \sum\limits_{i=1}^m B_i u_{i_t})\Delta t}\cdot \mathbf{z}_t \\\nonumber & \rightarrow & \mf{x}_{t + \Delta t} \approx \mf{\Psi}_d(\mf{z}_{t + \Delta t})=\hat{\mf{x}}_{t + \Delta t},   
\enl
assuming $\mf{u}_t$ stays almost constant over $\Delta t$. Traditionally, since we are mostly interested in future prediction of the state, KAE is trained by minimizing the difference between $\hat{\mf{x}}_{t_0 + (n+k) \Delta t}$ and $\mf{x}_{t_0 + (n+k) \Delta t}$, with $n,k\in\mb{Z}_+\cup \{0\}$. The loss functions traditionally used for training KAEs are the identity loss $\mc{L}_\text{id}$ ($k=0$), and the forward loss $L_\text{fwd}$ ($k=1,2,\ldots$) \cite{azencot2020forecasting}.   
\bnl\label{eq:KAE_loss}
    \mc{L}_{\text{id}} &=& \frac{1}{2M}\sum_{n=1}^M||\hat{\mathbf{x}}_{t_0 + n\Delta t}-\mathbf{x}_{t_0 + n\Delta t}||_2^2, \\ \nonumber \mc{L}_{\text{fwd}} &=& \frac{1}{2k_mM}\sum_{k=1}^{k_m}\sum_{n=1}^M||\hat{\mathbf{x}}_{t_0 + (n+k) \Delta t}-\mathbf{x}_{t_0 + (n+k) \Delta t}||_2^2,
\enl
where $||\cdot||_2$ is the 2-norm, $M$ is the number of samples over which we want to enforce the loss, and $k_m$ is the maximum value of $k$, i.e. the maximum look-ahead step for multi-step training.

\begin{figure}[t]
\centering 
\includegraphics[trim=0cm 0cm 0cm 0cm, clip=true, width=0.45\textwidth]{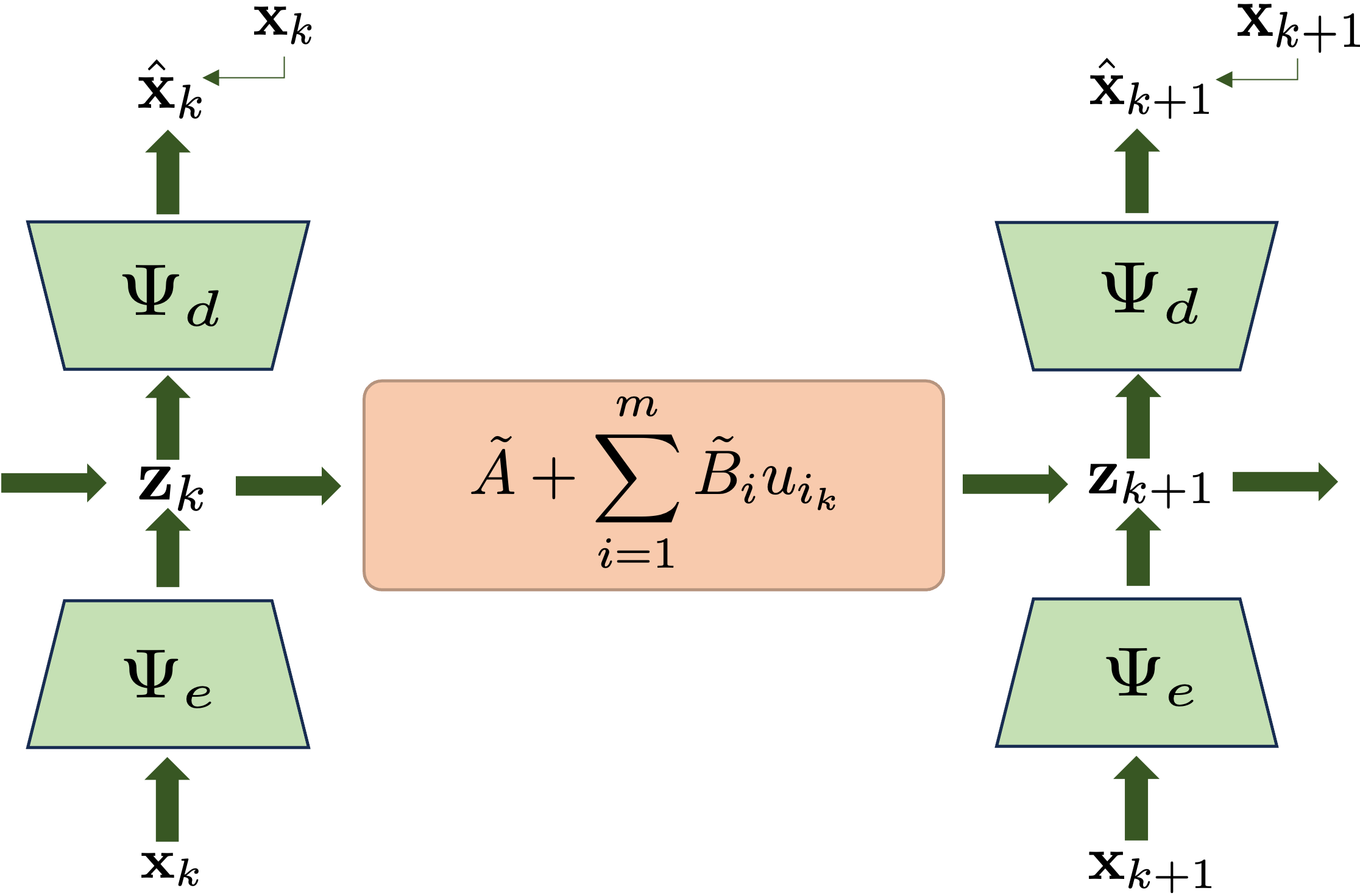}
\caption{Schematic of BLRAN (subscript $k$ denotes the time-sample at $t_0 + k\Delta t$)} \label{Fig: BLRAN}
\end{figure}

However, the KAE for control-affine system \eqref{eq:Control_KAE_operation} has a \emph{recurrent} structure which is exponential in control $\mf{u}_t$ and linear in state $\mf{z}_t$. We use a first-order Euler discretization to maintain the bilinear recurrent structure in the KAE as follows:
\bnl \label{eq:BLRAN}
\mathbf{z}_{t+\Delta t} &=& e^{(A +  \sum\limits_{i=1}^m B_i u_{i_t})\Delta t}\cdot \mathbf{z}_t\\\nonumber 
&\approx & (I_{N_l\times N_l} + A +  \sum\limits_{i=1}^m B_i u_{i_t})\mathbf{z}_t \Delta t \\\nonumber
& \approx & \left(\Tilde{A} + \sum\limits_{i=1}^m\Tilde{B_i} u_{i_t}\right) \mathbf{z}_t,
\enl
where $\Tilde{A} = (I_{N_l\times N_l} + A)\Delta t$ and $\Tilde{B_i} = B_i\Delta t$. We call this autoencoder structure (Fig. \ref{Fig: BLRAN}) for control-affine system bilinearly recurrent autoencoder (BLRAN). While this architecture has not been formally defined in any previous literature, it has been applied for quadrotor trajectory tracking \cite{Folkestad}.

\section{Temporally-Consistent Bilinearly Recurrent Autoencoders}
\begin{figure*}[t]
\centering 
\includegraphics[trim=0cm 0.5cm 0cm 0cm, clip=true, width=0.75\textwidth]{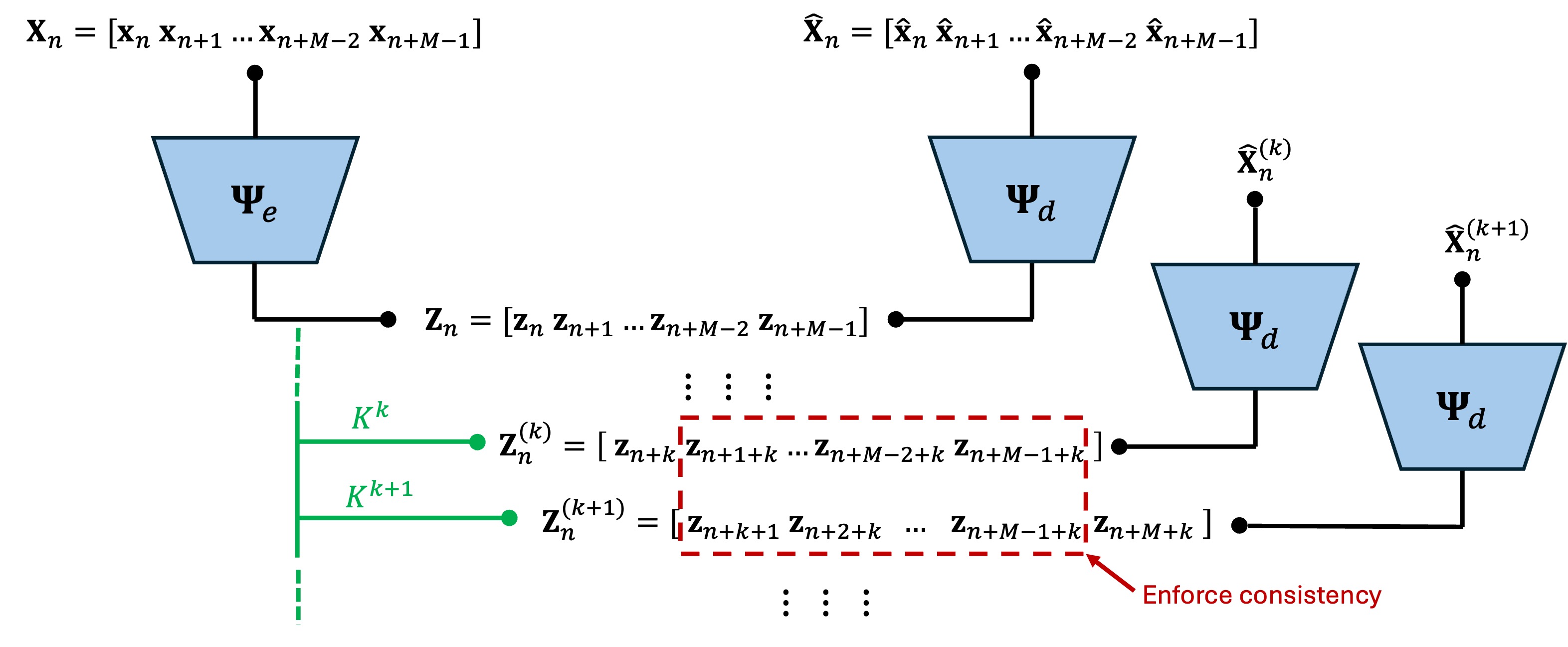}
\caption{Schematic of temporal consistency regularization (subscript $k$ denotes the time-sample at $t_0 + k\Delta t$)} \label{fig:tcBLRAN}
\end{figure*}
This section describes our proposed temporally-consistent bilinearly recurrent autoencoder (tcBLRAN) architecture. The key idea is to introduce a consistency regularization term for training BLRAN as described for KAE in \cite{Nayak2024}. The fundamental idea derives from the time-invariance property of the control-affine systems. In the context of \eqref{Eq: System}, the prediction consistency can be stated as the direct result of the following:
\bnl
   && \mf{x}_{t_n}+\int\limits_{t_n}^{t_n + k \Delta t} \left(\mf{f}(\mf{x}_t) + \sum\limits_{i=1}^m \mf{g}_i(\mf{x}_t) u_{i_t} \right)dt \\ &= & \mf{x}_{t_n - k\Delta t} + \int\limits_{t_n - k \Delta t}^{t_n  + k \Delta t}\left(\mf{f}(\mf{x}_t) + \sum\limits_{i=1}^m \mf{g}_i (\mf{x}_t) u_{i_t} \right)dt,\nonumber
\enl
$\forall n,k \in \mb{Z}_+\cup \{0\},~~k=1,2,\ldots, n, ~\text{and} ~\mf{x}_{t_n} = \mf{x}_{t_0 +n\Delta t}$. One key aspect of our approach is that instead of enforcing this consistency in the original state space, we do so in the Koopman invariant subspace (latent space). This helps avoid computation in the original high-dimensional space of nonlinear system, as well as provide robustness to noise. We justify enforcing such consistency in the latent space by summarizing the desired property of a consistent autoencoder $\mf{\Psi}_e$ and constructing the architecture of this framework. The autoencoder $\mf{\Psi}_e(\cdot) = [\psi_1(\cdot),\ldots,\psi_{N_l}(\cdot)]$ defines a vector valued observable of the state space $\mb{X}$ where each $\psi_i$ is a scalar valued function. Let $\mc{G}$ be the span of $\{\psi_1,\ldots,\psi_{N_l}\}$. Lemma \ref{Lem: tcBLRAN} summarizes the temporal consistency result under Assumption \ref{Assumption: Piecewise-constant control}.

\begin{assumption} \label{Assumption: Piecewise-constant control}
The control input is assumed to be piecewise-constant, i.e., $\mf{u}_t$ stays constant for a sampling interval $\Delta t$ satisfying $\mf{u}_t = \mf{u}_n$ for $t\in[t_0 + n\Delta t, t_0 + (n+1)\Delta t)$.
\end{assumption}

\begin{lemma} \label{Lem: tcBLRAN}
Let $\mc{G} = \operatorname{span} \{\psi_1,\ldots,\psi_{N_l}\}$ be invariant under $L_{\mf{f}}$ and $L_{\mf{g}_i}$, $i=1,\ldots,m$. Let the state and control trajectories $\mf{x}_t, \mf{u}_t$ be sampled at time instances $\{t_0, t_0 + \Delta t, \ldots, t_0 + n\Delta t, \ldots\}$. Then for a piecewise-constant control $\mf{u}_t$ as defined in Assumption \ref{Assumption: Piecewise-constant control}, the following is true:
\bnl \label{Eq: Lemma_tcBLRAN}
\mf{z}_t &=& \mf{\Psi}_e(\mf{x}_t) \\\nonumber
\mf{z}_{t_0 + (n + k) \Delta t} &=& \prod\limits_{\kappa=1}^{k} e^{(A +  \sum\limits_{i=1}^m B_i u_{i_{n+\kappa-1}})\Delta t} \mf{z}_{t_0 + n\Delta t},
\enl 
for all $n,k >0$ and the product is carried out from right to left.
\end{lemma}

\begin{proof}
Since $\mc{G}$ is invariant under $L_{\mf{f}}$ and $L_{\mf{g}_i}$, $i=1,\ldots,m$, then the transformation $\mf{z}_t = \mf{\Psi}_e(\mf{x}_t)$ satisfies the bilinear ODE:
\bql 
\dot{\mf{z}}_t = A\mf{z}_t + \sum\limits_{i=1}^m B_i\mf{z}_tu_{i_t},
\eql
with some $A, B_i,\, i\in\{1,\ldots,m\}$. Now, if the control input $\mf{u}_t,\, t\geq t_0$ is piecewise constant as described in Assumption \ref{Assumption: Piecewise-constant control}, then, for the resultant latent-state trajectory $\mf{z}_t$ satisfies
\[\mf{z}_{t_0 + (n + 1) \Delta t} =  e^{(A +  \sum\limits_{i=1}^m B_i u_{i_{n}})\Delta t} \mf{z}_{t_0 + n\Delta t},\]
for any time instance $t_0 + n\Delta t$. Now repeating this for $k$ times we get the desired result. 
\end{proof}

\begin{remark}
    The result in \eqref{Eq: Lemma_tcBLRAN} can be approximated by first order Euler discretization as described in \eqref{eq:BLRAN} to yield  
    \[ \mf{z}_{t_0 + (n + k) \Delta t} \approx  \prod\limits_{\kappa=1}^{k} \left(\Tilde{A} +  \sum\limits_{i=1}^m \Tilde{B}_i u_{i_{n+\kappa-1}}\right) \mf{z}_{t_0 + n\Delta t} \]
\end{remark}
Lemma \ref{Lem: tcBLRAN} is essentially heart of the temporal consistency loss. In order to define our loss function, let us consider the reference time-step to be $n$. For a batch size of $M$ (in latent space $\mf{Z}_n=[\mf{z}_{t_0 + n\Delta t} \ldots \mf{z}_{t_0 + (n+M-1)\Delta t}]$), let the BLRAN advance the dynamics up to $k_{tm}$ time-steps, giving rise to a set of batches $\mf{Z}^{(k)}_n=[\mf{z}_{t_0 +(n+k)\Delta t}^{(k)}~\mf{z}_{t_0 + (n+1+k)\Delta t}^{(k)}\ldots~\mf{z}_{t_0 +(n+M+k-1)\Delta t}^{(k)}]$ with $k = 1,2,\ldots,k_{tm}$, where the superscript $(k)$ denotes the multiplication with $\prod\limits_{\kappa=1}^{k} \left(\Tilde{A} +  \sum\limits_{i=1}^m \Tilde{B}_i u_{i_\kappa}\right)$ . The consistency loss $\mc{L}_\text{tc}$ (Fig. \ref{fig:tcBLRAN}) can be defined as,
\bnl\scriptsize\label{Eq: pc_loss}
    \mc{L}_\text{tc} &=& \frac{1}{2(k_{tm}-1)}\sum_{q=1}^{k_{tm} - 1}\mc{L}_q, \\\nonumber
    \mc{L}_{q} &=& \frac{1}{(k_{tm}-q)}\sum_{k=1}^{k_{tm}-q}\mc{L}_\text{k},\\\nonumber
    \mc{L}_\text{k} &=& \frac{1}{(M-q)}\sum_{p=q}^{M-1}||\mf{z}^{(k )}_{t_0 + (n+k+p)\Delta t}-\mf{z}^{(k + q)}_{t_0+(n+k+p)\Delta t}||_2^2
\enl
The total loss $(\mc{L}_\text{tot})$ for training tcBLRAN is given by,
\bnl
    \mc{L}_\text{tot}=\gamma_\text{id}\mc{L}_\text{id} + \gamma_\text{fwd} \mc{L}_\text{fwd} +  \gamma_\text{tc}\mc{L}_\text{tc},
\enl
Where $\gamma_\text{id}, \gamma_\text{fwd},$ and $\gamma_\text{tc}$ are the corresponding scaling factors.

% \textcolor{blue}{
% \begin{remark}
%     While other KAE architectures such as \cite{azencot2020forecasting} have been developed to enforce consistency for modeling autonomous systems, they rely on the time-reversible nature of dynamical systems. However, control systems are not time reversible, and therefore such consistency measures cannot be applied here. Enforcing temporal consistency, on the other hand, does not require any such assumptions, making it a meaningful framework to maintain consistency in the latent space of control-affine systems and ensure robust predictions across time.
% \end{remark}
% }

\begin{remark}
    Temporal consistency loss $\mc{L}_{\text{tc}}$ in \eqref{Eq: pc_loss} differs from the multi-step look-ahead prediction loss $\mc{L}_{\text{fwd}}$ in \eqref{eq:KAE_loss} because the former compares among predictions from different initial conditions to same final time $k_m$ rather than comparing the predictions to different final time with the \emph{labelled data}. Temporal consistency loss reduces BLRAN's dependence on large set of labelled training data.
\end{remark}

\begin{remark}
     While other KAE architectures \cite{azencot2020forecasting} enforced consistency by exploiting the the time-reversible nature of autonomous dynamical systems, it fails for control systems due to its time-irreversible nature. Temporal consistency loss, however, does not require time-reversibility, thereby making it practical for maintaining consistency in KAE architecture for control systems.
\end{remark}

\begin{figure}[t]
\centering 
\subfloat[]{\includegraphics[trim=1cm 7cm 0cm 7cm, clip=true, width=0.25\textwidth]{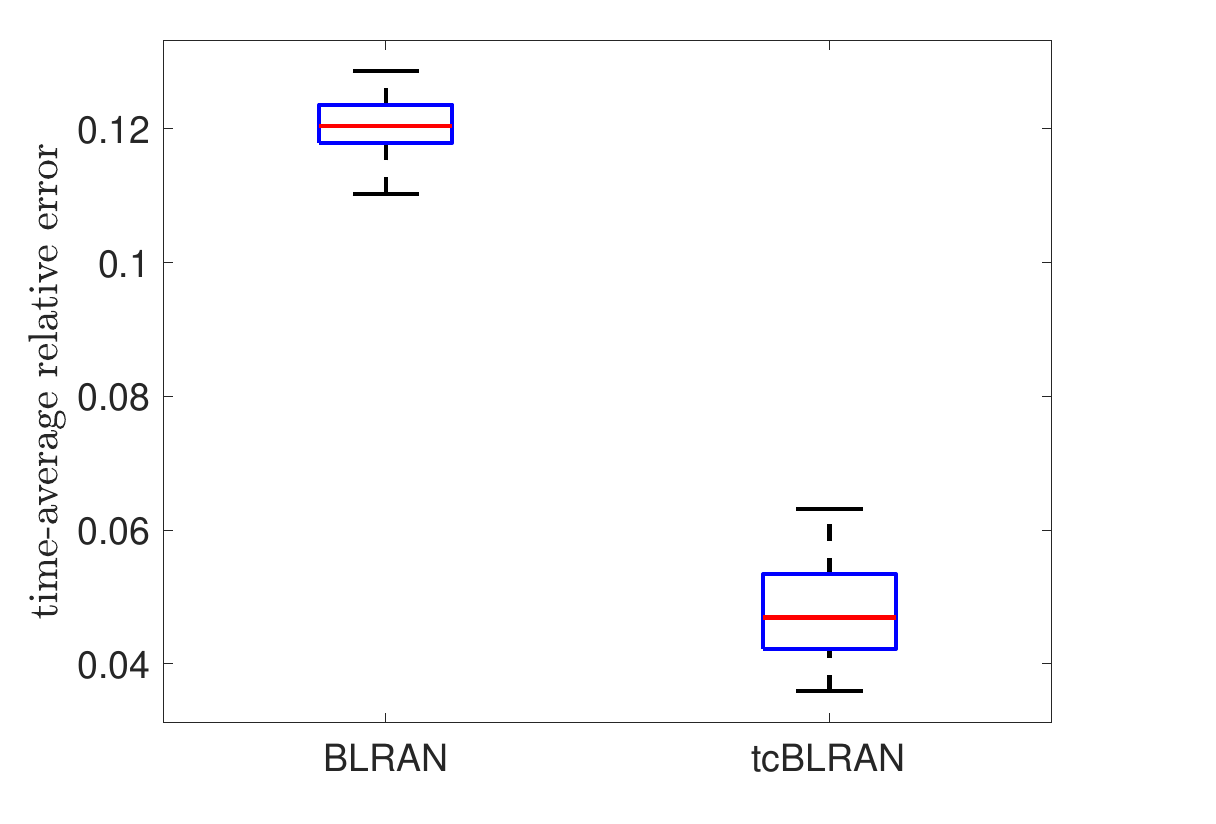}}
\subfloat[]{\includegraphics[trim=1cm 7cm 0cm 7cm, clip=true, width=0.25\textwidth]{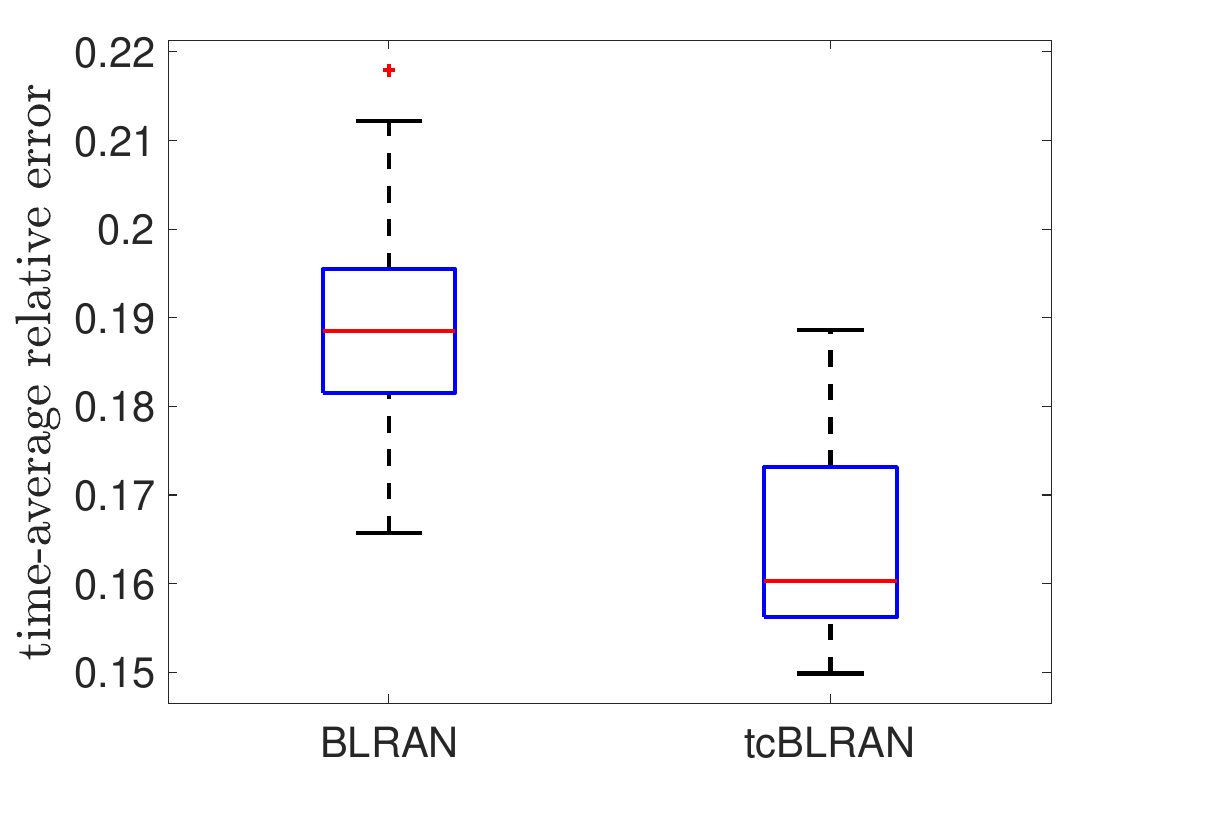}}
\caption{Time-average relative prediction error for simple pendulum \eqref{Eq: Pendulum} over 30 initial conditions and 10 seeds: (a) Trained on clean data, (b) Trained on noisy data with $20$ dB SNR.} \label{Fig: Error_Pendulum}
\end{figure}
\begin{figure}[!t]
\centering 
\includegraphics[trim=1cm 7cm 0cm 7cm, clip=true, width=0.4\textwidth]{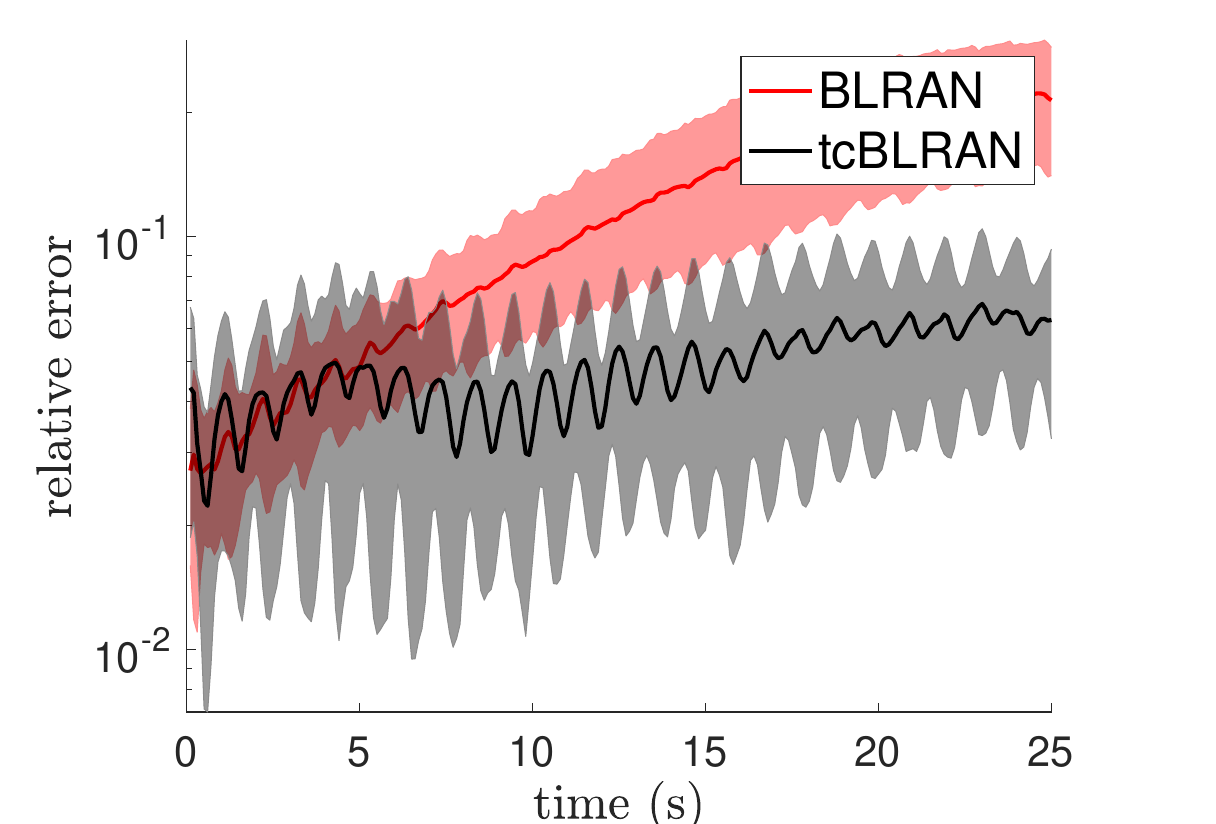}
\caption{Relative prediction error with time for simple pendulum \eqref{Eq: Pendulum}} \label{Fig: Time_error_pendulum}
\end{figure}

\section{Numerical Examples}
The effectiveness tcBLRAN is demonstrated on three benchmark problems: the simple pendulum, Van der Pol oscillator, and forced Duffing oscillator. All three systems were simulated over the time interval $t = [0, 220]$ with a sampling interval of $\Delta t = 0.1 s$, yielding 2200 equally spaced points. To mimic a high dimensional system, a random orthogonal transformation \cite{azencot2020forecasting} is used to rotate each system to a higher dimensional space of 64 dimensions. Each system is simulated from initial condition $[0.8, ~0]^{\top}$, and the time-average relative prediction error is calculated by simulating each system from 30 different initial conditions (taken as the last 30 points of the training dataset), with the control strategy employed being a piecewise-constant function with random scalar values in the range $[-0.15, ~0.15]$. The relative prediction error is calculated over $T = 25 s$ after rotating the state back to original dimension and is defined by
\bql
\text{error} = \frac{1}{T}\sum_{k=0}^{T-1} \frac{\|\hat{\mf{x}}_{t_0 + k\Delta t} - \mf{x}_{t_0 + k\Delta t} \|}{\|\mf{x}_{t_0 + k\Delta t}\|}.
\eql
To assess the tcBLRAN's performance under different noise conditions, the networks for each system are trained with both clean data and noisy data with $20$ dB SNR. In all three cases, tcBLRAN achieves superior prediction accuracy over the state of the art BLRAN for arbitrary piecewise constant control input. The tcBLRAN and BLRAN are trained and tuned over 10 different seeds for 600 epochs, and the learning rate decay is scheduled at epochs $[30, 100, 200, 400]$. Table \ref{tb:tcBLRANhyparam} contains the training details.

%%%%%%%%%%%%%%%%%%%%%%%%%%%%%%%%%%%%%

%I also deleted the relative prediction error vs time plots except for pendulum and changed the subsequent sections Simple Pendulum, Van der Pol oscillator and Forced Duffing oscillator slightly to reflect this.

%%%%%%%%%%%%%%%%%%%%%%%%%%%%%%%%%%%%%%

\begin{table*}[t]
\small
%\vspace{5pt}
\caption{tcBLRAN (BLRAN) hyperparameters: hyphen(-) indicates same value}\label{tb:tcBLRANhyparam}
\begin{center}
\begin{tabular}{rcccccc}
\hline
\qquad \qquad Hyperparameter \qquad \qquad \qquad \qquad \qquad\qquad \quad &  &  & Value \qquad\qquad \qquad \qquad\quad &  &  &\\
\end{tabular}
\begin{tabular}{rcccccc}
& \qquad \qquad \qquad \qquad \qquad  \qquad \qquad Pendulum \eqref{Eq: Pendulum} & & \qquad Van der Pol \eqref{Eq: VDP} & & \qquad Duffing \eqref{Eq: Duffing} & \\
 \end{tabular}
\begin{tabular}{rcccccc}
 SNR & Clean & 20 dB & Clean & 20 dB & Clean & 20 dB \\\hline
Learning rate  & 0.01 (-)   & 0.01 (-)   & 0.01 (-) & 0.01 (-) & 0.01 (-)  & 0.01 (-) \\
Learning rate decay & 0.5 (-) & 0.5 (-)  & 0.5 (-)  & 0.5 (-)  & 0.5 (-) & 0.5 (-) \\
Weight decay & 0.1 (0.01) & 1 (-)  & 1 (-)  & 1  (-) & 0.01 (0.1) & 1 (-) \\
Gradient clipping & 0.05 (-)  & 0.05 (-)  & 0.05 (-)  & 0.05 (-) & 0.05 (-) & 0.05 (-) \\
Look-ahead step $k_m$ & 12 (-)  & 12 (-)  & 32 (-)  & 32 (-)  & 12 (-) & 12 (-) \\
Temporal consistency step $k_{tm}$ & 2 (N/A)  & 2 (N/A)  & 2 (N/A)  & 2 (N/A)  & 2 (N/A) & 2 (N/A) \\
Batch size & 32 (-)  & 32 (-)  & 64 (-)  & 64 (-)  & 32 (-) & 32 (-)\\
$\gamma_{id}$  & 1 (-)  & 1 (-)  & 1 (-)  & 1 (-)  & 1 (-)  & 1 (-) \\
$\gamma_{fwd}$  & 1 (2)    & 0.5 (-)    & 1 (-)   & 2 (1) & 2 (-) & 2 (-)\\
$\gamma_{tc}$  & 2 (N/A)  & 0.5 (N/A)  & 0.01 (N/A) & 2 (N/A) & 0.5 (N/A) & 0.5 (N/A) \\
Latent state dimension $N_l$ & 12 (-) & 12 (-) & 12 (-) & 12 (-) & 12 (-) & 12 (-) \\
\# Nodes $\mf{\Psi}_e(\cdot)$ & 128 (-)  & 128 (-)  & 192 (-)  & 192 (-)  & 128 (-) & 128 (-)\\
\# Nodes $\mf{\Psi}_d(\cdot)$ & 128 (-)  & 128 (-)  & 192 (-)  & 192 (-)  & 128 (-) & 128 (-) \\
\hline 
\end{tabular}
\end{center}
\end{table*}
%
%
% \begin{table}[t]
% \vspace{5pt}
% \footnotesize
% \caption{BLRAN hyperparameters {\color{blue} Add}}\label{tb:BLRANhyparam}
% \begin{center}
% \begin{tabular}{rcc}
% \hline
%  \quad \qquad Hyperparameter & \qquad \qquad Value &  \\
%  & Pendulum \eqref{Eq: Pendulum} & Van der Pol \eqref{Eq: VDP}  \\
%  \end{tabular}
% \begin{tabular}{rcccc}
%  SNR & Clean & 20 dB & Clean & 20 dB \\\hline
% Time step $\Delta t$ & $0.1$s & $0.1$s & $0.1$s & $0.1$s\\
% Training length $N_{train}$ & $32$ & $32$s & $256$s & $256$s\\
% Learning rate  & 0.01   & 0.01   & 0.01  & 0.01\\
% Learning rate decay & 0.5 & 0.5 & 0.5 & 0.5 \\
% Weight decay & 0.01 & 1 & 1 & 1 \\
% Gradient clipping & 0.05 & 0.05 & 0.05 & 0.05 \\
% Look-ahead step $k_m$ & 12 & 12 & 32 & 32 \\
% Consistency step $p$ & 2 & 2 & 2 & 2 \\
% Batch size & 32 & 32 & 64 & 64 \\
% $\gamma_{id}$  & 1 & 1 & 1 & 1\\
% $\gamma_{fwd}$  & 1   & 0.5   & 1  & 1 \\
% \# $\mf{\Psi}_e(\cdot)$ layers & 3 & 3 & 3 & 3\\
% \# $\mf{\Psi}_d(\cdot)$ layers & 3 & 3 & 3 & 3\\
% \# Nodes $\mf{\Psi}_e(\cdot)$ & 128 & 128 & 192 & 192\\
% \# Nodes $\mf{\Psi}_d(\cdot)$ & 128 & 128 & 192 & 192\\
% \end{tabular}
% \end{center}
% \end{table}
%
\subsection{Simple pendulum}
The control-affine simple pendulum system, a canonical example in nonlinear dynamics, is modeled with the nonlinear equation 
\bql \label{Eq: Pendulum} \ddot{\theta} + \frac{g}{l} \sin(\theta) = u, \eql where \(\theta\) denotes the pendulum's angular displacement, \(g=9.8\) the gravitational acceleration, \(l=1\) the length of the pendulum, and \(u\) represents the control input. This is a two dimensional system with the state $[\theta \quad \dot\theta]^\top$.

%The control strategy employed is a piecewise-constant function, with random scalar values in the range of \([-0.15, 0.15]\).

\begin{figure}[t]
\centering 
\subfloat[]{\includegraphics[trim=1cm 7cm 0cm 7cm, clip=true, width=0.25\textwidth]{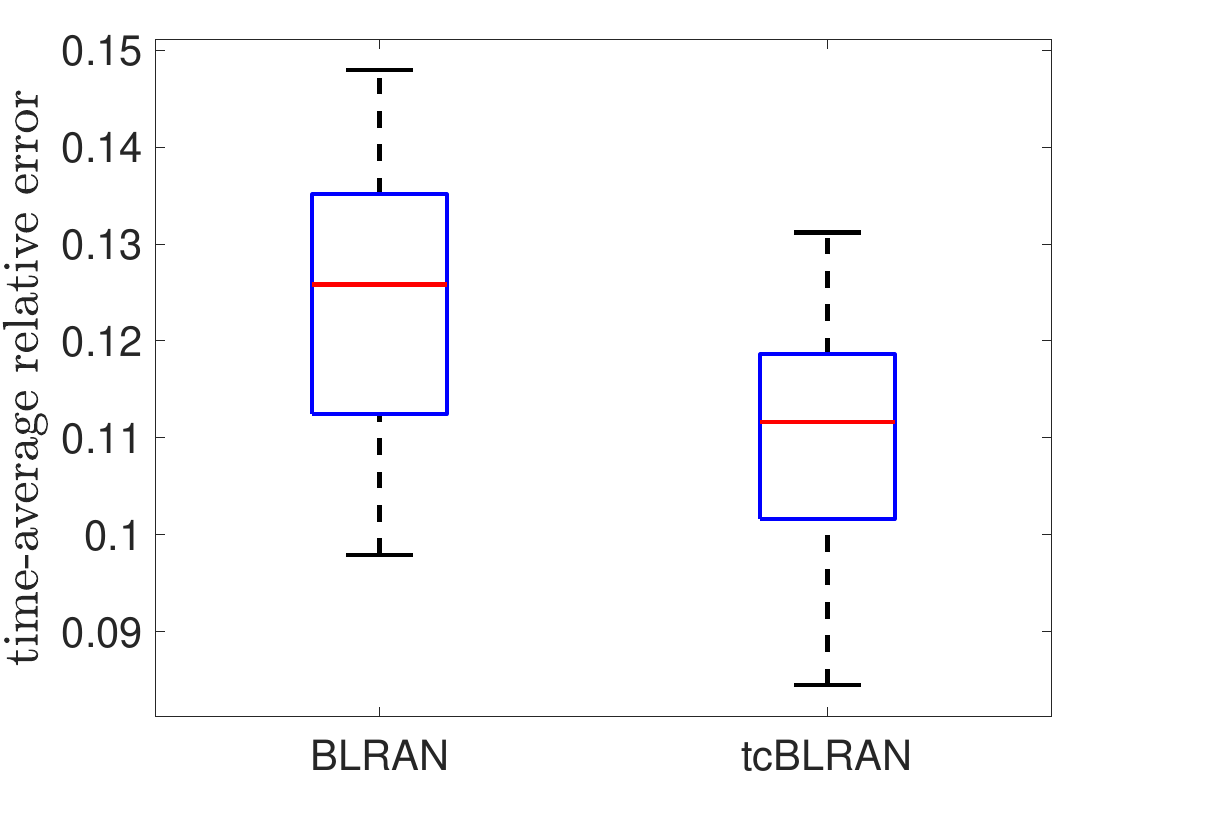}}
\subfloat[]{\includegraphics[trim=1cm 7cm 0cm 7cm, clip=true, width=0.25\textwidth]{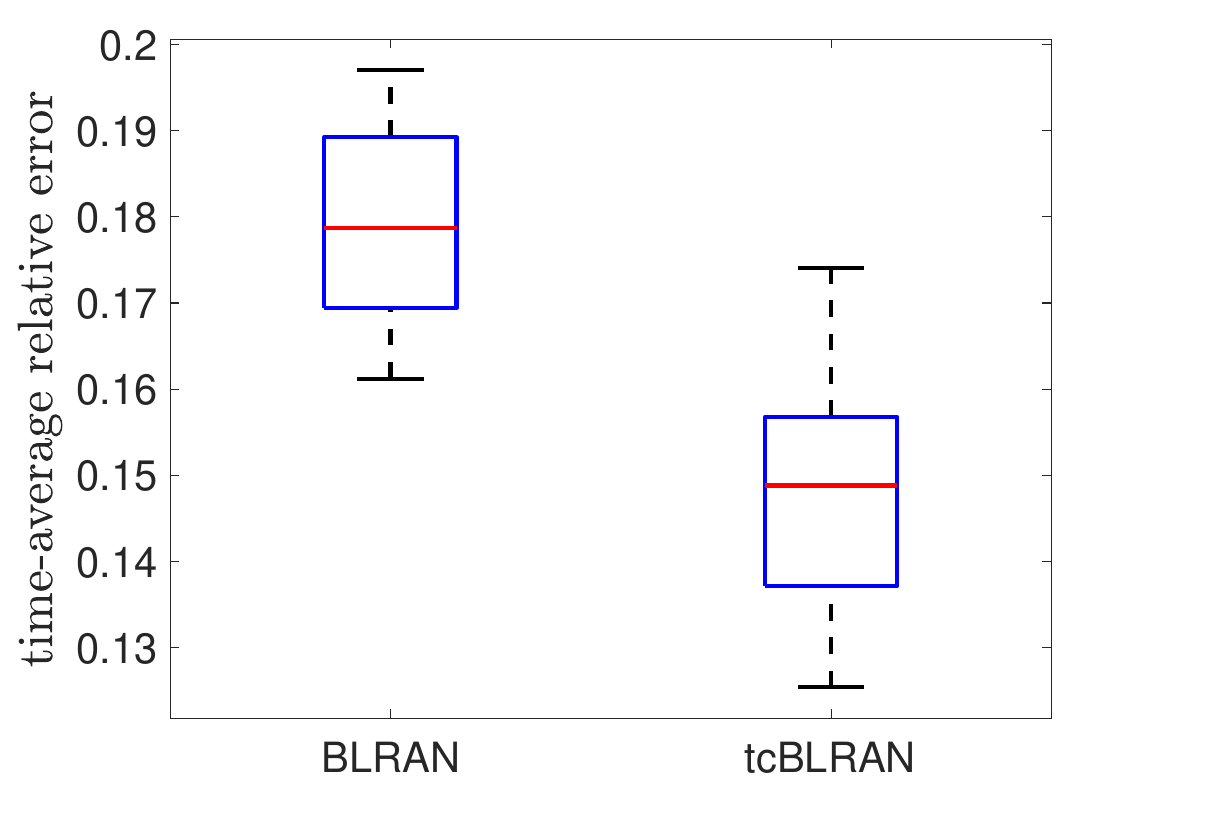}}
\caption{Time-average relative prediction error for Van der Pol oscillator \eqref{Eq: VDP} over 30 initial conditions and 10 seeds: (a) Trained on clean data, (b) Trained on noisy data with $20$ dB SNR.} \label{Fig: Error_VDP}
\end{figure}
%

\begin{comment}
\begin{figure}[!t]
\centering 
\subfloat[]{\includegraphics[trim=1cm 7cm 0cm 7cm, clip=true, width=0.25\textwidth]{CleanVDPTime.pdf}}
\subfloat[]{\includegraphics[trim=1cm 7cm 0cm 7cm, clip=true, width=0.25\textwidth]{SNR20VDPTime.pdf}}
\caption{Relative prediction error with time for Van der Pol oscillator \eqref{Eq: VDP}: (a) Trained on clean data, (b) Trained on noisy data with $20$ dB SNR.} \label{Fig: Time_error_VDP}
\end{figure}
\end{comment}

%The system was simulated over the time interval $t = [0, 220]$, with a 0.1-second sampling interval, yielding 2200 equally spaced data points.

%The system was initialized at an angle \(\theta_0 = 0.8\) with no initial angular velocity. To assess the tcBLRAN's performance under different noise conditions, the networks are trained for both the clean and noisy data with $20$ dB SNR.

Time-averaged relative prediction error for 30 different initial conditions are depicted in Fig.~\ref{Fig: Error_Pendulum} for clean and noisy data. Relative prediction error over $25$s  for clean training data of training length $N_{train}=32$ is shown in Fig.~\ref{Fig: Time_error_pendulum}.

% Table of prediction errors
%\begin{table}[t]
%\centering
%\caption{Results summary for simple pendulum. The quantity inside the paranthesis is the width of $90\%$ confidence interval}
%\label{tab:prediction_errors}
%\begin{tabular}{rcccc}
%{Model} & {$N_{train}$} & {Noise (SNR)} & {Prediction error (\%)} %\\ 
%BLRAN     & 32        & Clean      & x (y)     \\
%tcBLRAN    & 32        & 20      & x (y)       \\
%\end{tabular}
%\end{table}

\subsection{Van der Pol oscillator}
Next, tcBLRAN is applied to the controlled Van der Pol system \cite{Goswami2021}:
\bnl \label{Eq: VDP}
\ddot{x} = \mu(1-x^2)\dot{x}-x + u
% \dot{x} &=& v\nonumber\\
% \dot{v}&=& \mu(1-x^2)v-x + u.
\enl
The state $[x \quad \dot{x}]^\top$ is again two-dimensional. The damping coefficient $\mu$ is set to $1$. Time-average prediction errors are depicted in Fig.~\ref{Fig: Error_VDP} for training length $N_{train}=256$.

%The networks are again trained on both clean and noisy data.

\begin{figure}[t]
\centering 
\subfloat[]{\includegraphics[trim=1cm 7cm 0cm 7cm, clip=true, width=0.25\textwidth]{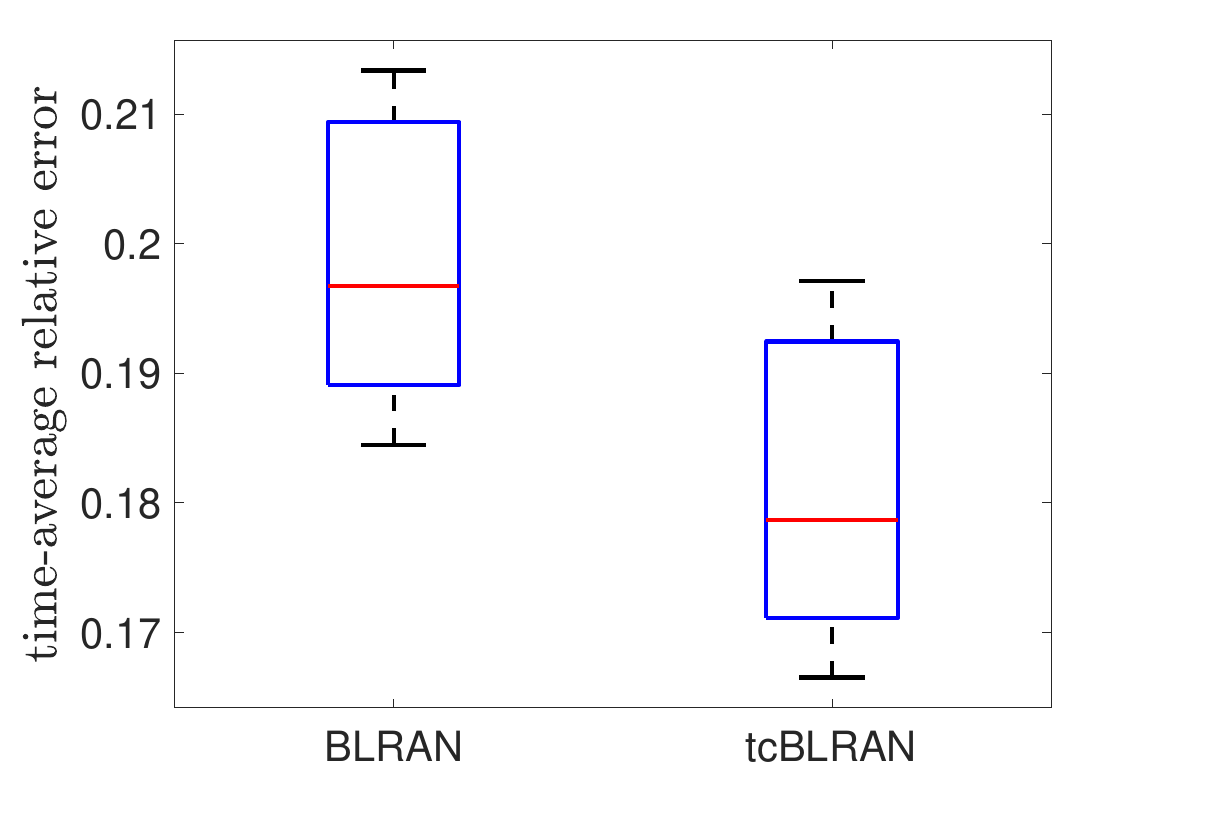}}
\subfloat[]{\includegraphics[trim=1cm 7cm 0cm 7cm, clip=true, width=0.25\textwidth]{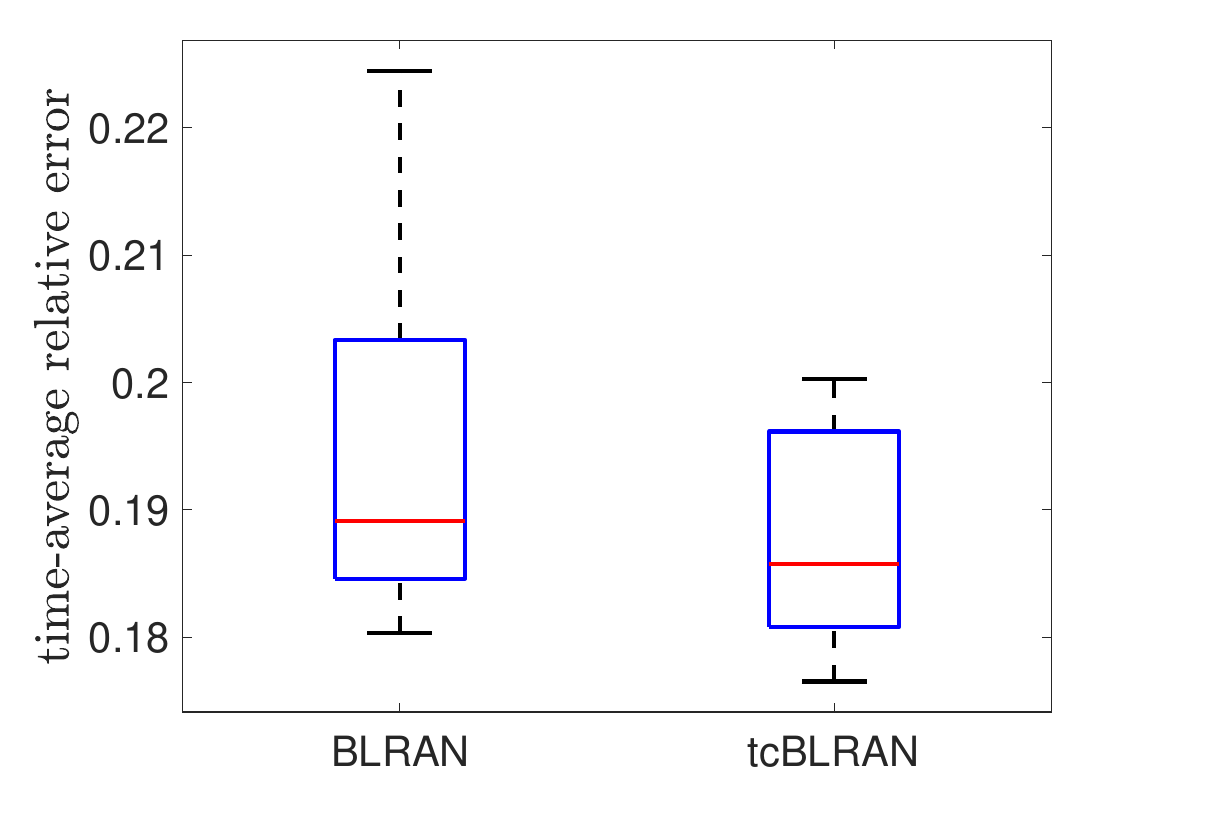}}
\caption{Time-average relative prediction error for forced Duffing oscillator \eqref{Eq: Duffing} over 30 initial conditions and 10 seeds: (a) Trained on clean data, (b) Trained on noisy data with $20$ dB SNR.} \label{Fig: Error_Duffing}
\end{figure}
%

\begin{comment}
\begin{figure}[!t]
\centering 
\subfloat[]{\includegraphics[trim=1cm 7cm 0cm 7cm, clip=true, width=0.25\textwidth]{CleanDuffingTime.pdf}}
\subfloat[]{\includegraphics[trim=1cm 7cm 0cm 7cm, clip=true, width=0.25\textwidth]{SNR20DuffingTime.pdf}}
\caption{Relative prediction error with time for forced Duffing oscillator \eqref{Eq: Duffing}: (a) Trained on clean data, (b) Trained on noisy data with $20$ dB SNR.} \label{Fig: Time_error_Duffing}
\end{figure}
\end{comment}

\subsection{Forced Duffing oscillator}
Lastly, tcBLRAN is used for modeling forced Duffing oscillator:
\bnl \label{Eq: Duffing}
\ddot{x} = -\delta \dot{x} - \alpha x - \beta x^3 + u
% \dot{x} &=& v\nonumber\\
% \dot{v} &=& -\delta v - \alpha x - \beta x^3 + u,
\enl
with $\alpha = -1$, $\beta=1$, and $\delta = 0.02$. Time-average prediction errors are shown in Fig.~\ref{Fig: Error_Duffing} with training length $N_{train}=32$.

%with control $\mf{u}_t\in [-0.15,\quad 0.15]$ and

\section{Conclusion}
We present temporally-consistent bilinearly recurrent autoencoder (tcBLRAN) algorithm for accurately modeling a control-affine nonlinear system from state and input data. Despite its advantages, tcBLRAN has some limitations. Temporal consistency introduces computational overhead which increases training time. Furthermore, with sufficient training data, the performance gap between tcBLRAN and BLRAN narrows significantly. These trade-offs highlight the necessity of tcBLRAN in scenarios with scarce or noisy data, where its robustness and enhanced prediction accuracy provide a critical edge. This method combines the Koopman bilinear form (KBF) and the time-invariance property of the control-affine system to enforce temporal consistency in the latent variables with a bilinear latent dynamics. The tcBLRAN is evaluated and compared against the vanilla BLRAN on three systems with clean and noisy data. It outperforms the vanilla autoencoder methods for more accurate prediction which is important for control systems application. In future, the tcBLRAN model will be used for data-driven model-predictive control (MPC) of nonlinear system.

\section*{Acknowledgement}
The authors acknowledge the support of Ohio Supercomputer Center (PAS2709) for the computational resources.
%\balance
\bibliographystyle{IEEEtran}
\bibliography{bibl}
\end{document}